\documentclass[
    ,final            
  ]
  {aipproc}

\layoutstyle{8x11double}

\begin{document}

\title{Guaranteed discovery of the NmSuGra model}


\author{Csaba Bal\'azs}{
  address={School of Physics, Monash University, Melbourne Victoria 3800, Australia}
}

\author{Daniel Carter}{
  address={School of Physics, Monash University, Melbourne Victoria 3800, Australia}
}

\begin{abstract}

We analyze the discovery potential of the next-to-minimal supergravity motivated 
model: NmSuGra.  This model is an extension of mSuGra by a gauge singlet, and 
contains only one additional parameter: $\lambda$, the Higgs-singlet-Higgs 
coupling.  NmSuGra solves the $\mu$-problem and reduces the fine tuning of mSuGra.
After identifying parameter space regions preferred by present experimental 
data, we show that these regions of NmSuGra are amenable to detection by 
the combination of the Large Hadron Collider and an upgraded Cryogenic Dark 
Matter Search. 
This conclusion holds strictly provided that the more than three sigma 
discrepancy in the difference of the experimental and the standard theoretical 
values of the anomalous magnetic moment of the muon prevails in the future.

\end{abstract}

\maketitle


\section{Introduction}
\label{sec:Introduction}


Supersymmetry is very successful in solving outstanding problems of the standard 
model (SM) of elementary particles, including the hierarchy of fundamental 
energy scales, the existence and properties of dark matter and the unification 
of gauge forces. 
%
%
However, the minimal supersymmetric standard model (MSSM) suffers from several 
problems.  For example, the $\mu$ term is not protected from radiative corrections, and its 
viable parameter regions are now quite fine tuned \cite{Giudice:2008bi}.  
Gauge-singlet extensions of the MSSM offer solutions to these problems.  In the 
next-to-minimal MSSM (NMSSM), the $\mu$ term is dynamically generated and no 
dimensionful parameters are introduced in the superpotential (other than the 
vacuum expectation values that are all naturally weak-scale), making the NMSSM a 
truly natural model (see references in \cite{Balazs:2008ph}).  We define the 
next-to-minimal supergravity motivated (NmSuGra) model, imposing universality of 
sparticle masses, gaugino masses, and trilinear couplings at the grand 
unification theory (GUT) scale.


Using a simple likelihood analysis, we first identify the parameter regions of 
the NmSuGra model that are preferred by present experimental data.  We 
combine theoretical exclusions with limits from the CERN Large Electron-Positron 
(LEP) collider, the Fermilab Tevatron, NASA's Wilkinson Microwave Anisotropy 
Probe (WMAP) satellite, the 
Soudan Cryogenic Dark Matter Search (CDMS), the Brookhaven Muon g$-$2 
Experiment, and various b-physics measurements including $b \to s \gamma$ and $B_s \to l^+ l^-$. 
%
%
We then show that, assuming recent results on the muon g$-$2 are accurate, the 
favored parameter space can be detected by the combination of the LHC and an 
upgraded CDMS. (See Ref.s \cite{Stockinger:2007pe, Passera:2008jk} on the 
theoretical uncertainty of $\Delta a_{\mu}$.)

\section{The NmSuGra model}


In this work, we adopt the superpotential 
\begin{eqnarray}
 W = W_{Y} + \lambda \hat{S} \hat{H}_u \cdot \hat{H}_d + \frac{\kappa}{3} \hat{S}^3,
\label{eq:W}
\end{eqnarray}
where $W_{Y}$ is the MSSM Yukawa superpotential, $\hat{S}$ ($\hat{H}_{u,d}$) is 
a standard gauge singlet ($SU(2)_L$ doublet) chiral superfield, $\lambda$ and 
$\kappa$ are dimensionless couplings, and $\hat{H}_u \cdot \hat{H}_d = 
\epsilon_{\alpha\beta} \hat{H}_u^\alpha \hat{H}_d^\beta$ with $\epsilon_{11} = 
1$.  The corresponding soft supersymmetry breaking terms are
\begin{eqnarray}
 \mathcal{L}^{soft} = \mathcal{L}^{soft}_{MSSM} + m_S^2 |S|^2 + \nonumber \\
 (\lambda A_\lambda S H_u \cdot H_d + \frac{\kappa A_\kappa}{3} S^3 + h.c.),
\end{eqnarray}
where $\mathcal{L}^{soft}_{MSSM}$ contains no $B \mu$ term%
\footnote{Radiative breaking the $Z_3$ symmetry may destabilize the hierarchy of 
vevs in the NMSSM, however by imposing a $Z_2$ R-symmetry these problems can be 
alleviated without affecting the phenomenology 
\cite{Panagiotakopoulos:1998yw}.}.


We assume that the soft masses of the gauginos unify to $M_{1/2}$, those of the 
sfermions and Higgses to $M_0$, and all the trilinear couplings (including 
$A_\kappa$ and $A_\lambda$) to $A_0$ at the GUT scale.  Defining $\mu = \lambda 
\langle S \rangle$, and $\tan\beta = \langle H_u \rangle/\langle H_d \rangle$ 
(the ratio of Higgs vevs), our free parameters are \cite{Balazs:2008ph}: 
\begin{eqnarray} 
M_0, ~ M_{1/2}, ~ A_0, ~ \tan\beta, ~ \lambda, ~ {\rm sign}(\mu). 
\label{eq:5Para} \end{eqnarray}
To keep all the attractive features of the CMSSM/ mSuGra, we adhere to 
universality and use only $\lambda$ to parametrize the singlet sector.  This 
minimal extension alleviates problems of CMSSM/mSuGra rooted in the MSSM.  Other 
constrained versions of the NMSSM have been studied in the recent literature 
\cite{Djouadi:2008yj, Hugonie:2007vd, Belanger:2005kh, Cerdeno:2007sn, 
Djouadi:2008uw} (see also the contributions by U. Ellwanger in these proceedings).


Our goal is to show that the NmSuGra model can be discovered by nascent 
experiments in the near future.  To this end, for each set of the model 
parameters, we quantify the experimental preference in terms of:
\begin{eqnarray}
\sqrt{\chi^2} = \biggl(\sum_{i=1}^7
\Bigl(\frac{m_i^{experiment}-m_i^{NmSuGra}}{\sigma_i}\Bigr)^2\biggr)^{1/2}
\label{eq:Chi2}
\end{eqnarray}
where $m_i$ is the central value of a physical quantity measured by an 
experiment or calculated in the NmSuGra model, and $\sigma_i$ is the combined 
experimental and theoretical uncertainty.  We include 
experimental upper limits for 
 $\Omega h^2 = 0.1143 \pm 0.0034$ \cite{Komatsu:2008hk}, 
 $Br(B_s \to \mu^+\mu^-) = 5.8 \times 10^{-8}$ (95 \% CL) \cite{Barberio:2006bi} and 
 $\sigma_{SI}$ by CDMS \cite{Ahmed:2008eu}.
We also include the LEP lower limits of the lightest scalar Higgs and chargino masses (which can 
be approximately stated as) \cite{Abbiendi:2003sc}:
 $m_h > 114.4$ GeV for $\tan\beta {\scriptstyle \stackrel{<}{\sim}} 10$,
      $m_h > 91$ GeV for $\tan\beta {\scriptstyle \stackrel{>}{\sim}} 10$,
 $m_{\tilde{W}_1} > 104$ GeV.
Finally, we consider the central values of 
 $\Delta a_\mu = 29.5 \pm 8.8 \times 10^{-10}$ \cite{Stockinger:2007pe}, and 
 $Br(b \to s \gamma) = 3.55 \pm 0.26 \times 10^{-4}$ \cite{Barberio:2006bi}.  
The related uncertainties are given above at 68 \% CL, unless stated otherwise.
Theoretical uncertainties are calculated using NMSSMTools \cite{Ellwanger:2006rn} 
for the b-physics related quantities.

\begin{figure}[t]
\includegraphics[width=0.48\textwidth,height=0.43\textwidth]{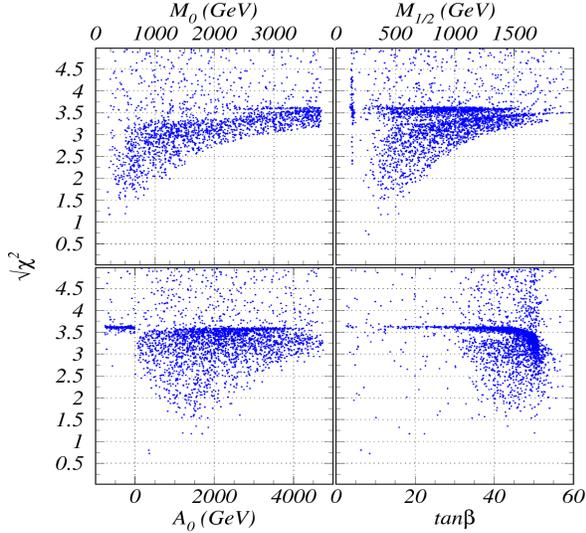}
\caption{\label{fig:Chi2VsInput} 
The log-likelihood vs. four of the NmSuGra 
input parameters for a random sample of models.  The combination of the 
experimental quantities included in $\chi^2$ favor low values of $M_0$, 
$M_{1/2}$ and $|A_0|$.
}
\end{figure}

A glance at the likelihood reveals a significant statistical preference for 
relatively narrow intervals of $M_0$, $M_{1/2}$ and $A_0$, as shown in figure 
\ref{fig:Chi2VsInput} for a randomly selected set of models with ${\rm 
sign}(\mu) > 0$.  At high values of $M_0$, $M_{1/2}$ and $|A_0|$, $\chi^2$ is 
dominated by $\Delta a_\mu$, similar to the CMSSM.  Based on this, we limit our 
study to the following ranges of the continuous parameters: $0 < M_0 < 4 ~{\rm 
TeV}, 0 < M_{1/2} < 2 ~{\rm TeV}, 0 < |A_0| < 5 ~{\rm TeV}, 1 < \tan\beta < 60, and
0.01 < \lambda < 0.7$.

\section{Detectability of NmSuGra}

\begin{figure}[t]
\includegraphics[width=0.48\textwidth,height=0.43\textwidth]{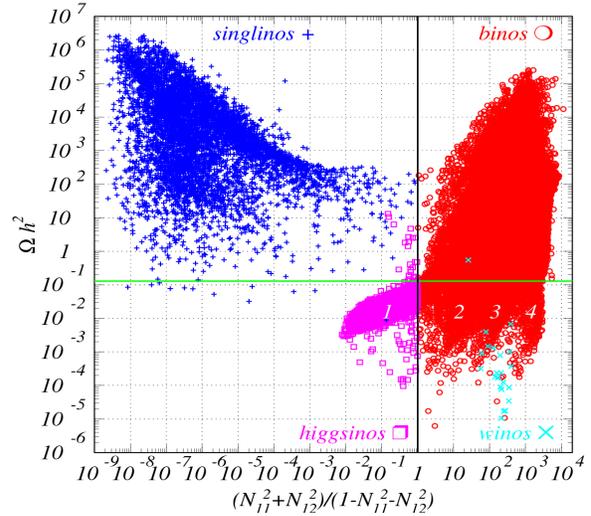}
\caption{\label{fig:Oh2vsAdMix} 
Relic abundance of the lightest neutralino as the function of its gaugino 
admixture.  Right of the vertical line the neutralino is mostly gaugino. The 
horizontal lines shows the WMAP upper limit (95 \% CL).
}
\end{figure}

Having defined the NmSuGra model and the range of its parameters, 
we set out to show that this parameter region will be detectable by the LHC and 
an upgraded CDMS detector.
Two million theoretically allowed representative model points (from an initial 
sample of 20 million) are projected in figure \ref{fig:Oh2vsAdMix} to the plane 
of $\Omega h^2$ vs. the gaugino admixture of the lightest neutralino. 
From figure \ref{fig:Oh2vsAdMix} it is evident that the WMAP upper limit (green 
horizontal line) favors models with mostly bino- (red circles) and higgsino-like 
(magenta squares) lightest neutralino, while the fraction of allowed models with 
singlino-like (blue pluses) dark matter is negligible.
By checking mass relations and couplings, we can easily establish that branch 4 
contains only models with dominant neutralino-stop coannihilation, while branch 
3 corresponds to neutralino-stau coannihilation. Branch 2 represents the Higgs 
resonance corridors, and branch 1 is the equivalent of the CMSSM/mSuGra focus 
point region. 


\begin{figure}[t]
\includegraphics[width=0.48\textwidth,height=0.43\textwidth]{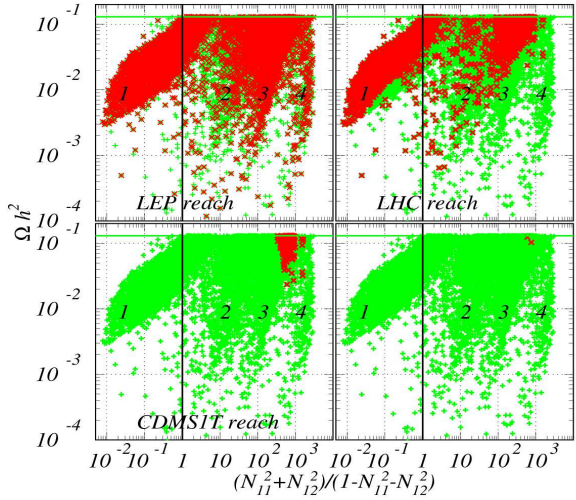}
\caption{\label{fig:Oh2vsAdMix5}
Same as figure \ref{fig:Oh2vsAdMix}, but green (plus) models can be reached by the 
combination of LEP, the LHC and CDMS1T, while red ones (crosses) cannot.  The last 
frame dismisses experimentally inaccessible points which have $\sqrt{\chi^2}>3$. 
}
\end{figure}

To gauge the detectability of the NmSuGra model, first we identify model points 
that could have been seen at LEP \cite{Balazs:2008ph}.  The top left frame of 
figure \ref{fig:Oh2vsAdMix5} shows model points from 
figure \ref{fig:Oh2vsAdMix} colored either green (plus) if the model would be 
accessible to LEP or red (cross) if it passes the above LEP constraints and is 
therefore allowed.  Just as in the CMSSM/mSuGra the neutralino-stop 
coannihilation region is mostly covered by LEP.
%
%
For the LHC reach we use a conservative approximation relying on the similarity 
between the mSuGra and NmSuGra models.  According to Ref. \cite{Baer:2003wx} the 
reach of the LHC for mSuGra can be well approximated by the combined reach for 
gluinos and squarks.  Based on this, if either the gluino mass is below 1.75 TeV, 
or the geometric mean of the stop masses is below 2 TeV for a given model point, 
we consider it discoverable at the LHC.


The top right frame of figure \ref{fig:Oh2vsAdMix5} shows the model points that 
can be reached by LEP and the LHC using the above criteria.  As in the 
CMSSM/mSuGra, most of the slepton coannihilation and the bulk of the Higgs 
resonance branches are covered by the LHC.  A good part of the focus point is 
also within reach of the LHC, with the exception of models with high $M_0$ 
and/or $M_{1/2}$.  
%
%
The bottom left frame of figure \ref{fig:Oh2vsAdMix5} shows the reach of a one ton 
equivalent of CDMS (CDMS1T).  As expected from the CMSSM/mSuGra, the rest of the 
focus point and most of the remaining Higgs resonances are in the reach 
of CDMS1T.  The small number of models that remain inaccessible are all located 
in regions that have relatively low $M_0$ and high $M_{1/2}$ with dominant 
neutralino annihilation via s-channel Higgs resonances.  The NmSuGra contribution 
to $\Delta a_\mu$ in these model points is outside the preferred 99 \% CL 
region as shown by the last frame.

Assuming that the NmSuGra contribution to the anomalous magnetic moment of the 
muon is larger than $3.1 \times 10^{-10}$ constrains slepton and chargino masses 
below 3 and 2.5 TeV, respectively.  Since universality restricts the mass 
hierarchy within NmSuGra, the resulting mass spectrum is typically mSuGra-like. 
Thus, the cascade decays and their signatures at LHC are not expected to deviate 
significantly from that of the mSuGra case.

\section{Conclusions}

Analyzing the next-to-minimal supergravity motivated (NmSuGra) model, we found 
that the LHC and an upgraded CDMS experiment will be able to discover the 
experimentally favored regions of this model provided that the present deviation 
between the experimental and standard theoretical values of the muon anomalous 
magnetic moment prevails.


\begin{theacknowledgments}
We thank M. Carena, U. Ellwanger, A. Menon, D. Morrissey, C. Munoz and C. Wagner 
for invaluable discussions on various aspects of the NMSSM.  This research was 
funded in part by the Australian Research Council under Project ID DP0877916.
\end{theacknowledgments}

\bibliographystyle{aipproc}

\bibliography{references}

\begin{thebibliography}{16}
\expandafter\ifx\csname natexlab\endcsname\relax\def\natexlab#1{#1}\fi
\providecommand{\enquote}[1]{``#1''}
\expandafter\ifx\csname url\endcsname\relax
  \def\url#1{\texttt{#1}}\fi
\expandafter\ifx\csname urlprefix\endcsname\relax\def\urlprefix{URL }\fi
\providecommand{\eprint}[2][]{\url{#2}}

\bibitem[Giudice(2008)]{Giudice:2008bi}
G.~F. Giudice  (2008), \eprint{0801.2562}.

\bibitem[Balazs and Carter(2008)]{Balazs:2008ph}
C.~Balazs, and D.~Carter  (2008), \eprint{0808.0770}.

\bibitem[Stockinger(2007)]{Stockinger:2007pe}
D.~Stockinger  (2007), \eprint{0710.2429}.

\bibitem[Passera et~al.(2008)]{Passera:2008jk}
M.~Passera, W.~J. Marciano, and A.~Sirlin  (2008), \eprint{0804.1142}.

\bibitem[Panagiotakopoulos and Tamvakis(1999)]{Panagiotakopoulos:1998yw}
C.~Panagiotakopoulos, and K.~Tamvakis, \emph{Phys. Lett.} \textbf{B446},
  224--227 (1999), \eprint{hep-ph/9809475}.

\bibitem[Djouadi et~al.(2008{\natexlab{a}})]{Djouadi:2008yj}
A.~Djouadi, U.~Ellwanger, and A.~M. Teixeira  (2008{\natexlab{a}}),
  \eprint{0803.0253}.

\bibitem[Hugonie et~al.(2007)]{Hugonie:2007vd}
C.~Hugonie, G.~Belanger, and A.~Pukhov, \emph{JCAP} \textbf{0711}, 009 (2007),
  \eprint{0707.0628}.

\bibitem[Belanger et~al.(2005)]{Belanger:2005kh}
G.~Belanger, F.~Boudjema, C.~Hugonie, A.~Pukhov, and A.~Semenov, \emph{JCAP}
  \textbf{0509}, 001 (2005), \eprint{hep-ph/0505142}.

\bibitem[Cerdeno et~al.(2007)]{Cerdeno:2007sn}
D.~G. Cerdeno, E.~Gabrielli, D.~E. Lopez-Fogliani, C.~Munoz, and A.~M.
  Teixeira, \emph{JCAP} \textbf{0706}, 008 (2007), \eprint{hep-ph/0701271}.

\bibitem[Djouadi et~al.(2008{\natexlab{b}})]{Djouadi:2008uw}
A.~Djouadi, et~al.  (2008{\natexlab{b}}), \eprint{0801.4321}.

\bibitem[Komatsu et~al.(2008)]{Komatsu:2008hk}
E.~Komatsu, et~al.  (2008), \eprint{0803.0547}.

\bibitem[Barberio et~al.(2006)]{Barberio:2006bi}
E.~Barberio, et~al.  (2006), \eprint{hep-ex/0603003}.

\bibitem[Ahmed et~al.(2008)]{Ahmed:2008eu}
Z.~Ahmed, et~al.  (2008), \eprint{0802.3530}.

\bibitem[Abbiendi et~al.(2004)]{Abbiendi:2003sc}
G.~Abbiendi, et~al., \emph{Eur. Phys. J.} \textbf{C35}, 1--20 (2004),
  \eprint{hep-ex/0401026}.

\bibitem[Ellwanger and Hugonie(2007)]{Ellwanger:2006rn}
U.~Ellwanger, and C.~Hugonie, \emph{Comput. Phys. Commun.} \textbf{177},
  399--407 (2007), \eprint{hep-ph/0612134}.

\bibitem[Baer et~al.(2003)]{Baer:2003wx}
H.~Baer, C.~Balazs, A.~Belyaev, T.~Krupovnickas, and X.~Tata, \emph{JHEP}
  \textbf{06}, 054 (2003), \eprint{hep-ph/0304303}.

\end{thebibliography}

\end{document}